# Particle trapping and beaming in a 3D nanotip excited with plasmonic vortex

Xue Jin Zhang,[1] Yuefeng Shen[1], Kai Liu[1], Xueyun Li[1], Nicolò Maccaferri[2], Yuri Gorodetski[3*] and Denis Garoli[4*]

[1]*National Laboratory of Solid State Microstructures and Collaborative Innovation Center of Advanced Microstructures and School of Physics and College of Engineering and Applied Sciences, Nanjing University, Nanjing 210093, China.*
[2]*Physics and Materials Science Research Unit, University of Luxembourg, 162a avenue de la Faïencerie, L-1511 Luxembourg, Luxembourg*
[3]*Electrical Engineering and Electronics Department, Mechanical Engineering and Mechatronics Department, Ariel University, Ariel 40700, Israel.*
[4]*Istituto Italiano di Tecnologia, Via Morego 30, 16163, Genova (Italy).*
*Corresponding authors: denis.garoli@iit.it; yurig@ariel.ec.il*

**Recent advances in nanotechnologies have prompted the need for tools to accurately and non-invasively manipulate individual nano-objects. Among the possible strategies, optical forces have been widely used to enable nano-optical tweezers capable of trapping or moving a specimen with unprecedented accuracy. Here, we propose an architecture consisting of a nanotip excited with a plasmonic vortex enabling effective dynamical control of nanoparticles in three dimensions. The optical field generated by the structure can be used to manipulate single dielectric nanoparticles acting on the total angular momentum of light used to illuminate the structure. We demonstrate that it's possible to stably trap or push the particle from specific points, thus enabling a new platform for nanoparticle manipulation and sorting.**

Optical trapping technologies find now application in several fields from biology to nanoscience [1]. Standard methods use a focused laser beam to efficiently trap and manipulate nano-objects with spatial resolution close to the wavelength of illumination. Subwavelength spatial resolution can be achieved with innovative approaches based on engineered metallic nanostructures. In particular, these can generate electromagnetic (EM) surface waves known as surface plasmon polaritons (SPPs) enabling the EM fields control and confinements down to a few nm [2]. The ability of SPPs to squeeze light into nanometer scale volumes with consequent enhancement in the local field intensity opened the way to several applications from photovoltaics and spectroscopy to imaging, and sensing [3]. Strong EM field confinement suggests the use of plasmonic nanostructures for trapping. Several works have demonstrated how plasmonics can offer significant advantages over conventional optical tweezers [1,4–9]. In most of the cases, plasmonic trapping and tweezing has been investigated in planar metallic nanostructures optimized for a specific wavelength. With these structures [9,10] it has been possible to experimentally demonstrate tweezing and manipulation of dielectric and metallic nanoparticles down to 20 nm in diameter. One of the most desired functionalities that essentially could be achieved with plasmonics is the possibility to dynamically trap and release nanoparticles, also in three dimensions [11], and to achieve real-time sorting and assembly [12]. Here we show that by tuning the polarization of the incident light, one can achieve such a dynamic plasmonic trapping which can be used to transport or rotate nanoparticles at locations determined by the pre-fabricated nanostructure on-chip [12]. Noteworthy, the polarization of the incident light can be utilized as an additional degree of freedom when considering an optical angular momentum (AM). The total amount of the AM per photon carried by a quasi-paraxial beam is given as (in units of ℏ) $j = l + s$; where $s$ is the spin orbital momentum (the handedness light polarization) and $l$ is the orbital angular momentum (OAM) [13]. The latter (usually referred as the topological charge) corresponds to the winding of spiral phase ( $\phi$ ) and can be derived as, $l = \frac{1}{2\pi} \oint d\phi$. Clearly, the azimuthal energy flow of optical beams with OAM can induce a rotational motion of nano-objects while the spin part is expected to spin the particles about their own axis [12,14]. While in free-propagating beams the two components of the AM are ideally uncoupled, the light-matter interaction resulting in SPP excitation was shown to lead to spin-orbit coupling. This provides the unique ability to manipulate the OAM by externally modifying the incident light polarization. For this reason, generation and manipulation of plasmonic distributions with OAM has been a subject of several recent works [15–17]. The SPPs that carry OAM have been referred as the plasmonic vortices (PVs). Due to the plasmonic confinement to the metal surface, the 3D OAM phenomena can be coupled to the inherently 2D geometry of the SPP waves. This phenomenon enables new ways of manipulating the EM fields. In particular, we've recently demonstrated that a 3D tapered cone can be used to couple a propagating PV to a far-field light beaming or efficiently focus it at the apex [15]. The investigated structure comprised of a conical nanotip grown on a metal surface with a coupler spiral grating around it. Here we show that such design can be utilized as a polarization dependent optical tweezer for the near-field particle manipulation. We use a Comsol software to simulate the particle behavior in the plasmonic field distributions induced by different incident polarization states. The calculated data shows that a properly designed structure may perform rather as an optical trap pinning a particle in the vicinity of the tip end or as a motor to push it to the far-field.

Figure 1 illustrates the investigated structure design and the two main configurations of excitations: focusing and beaming [15].

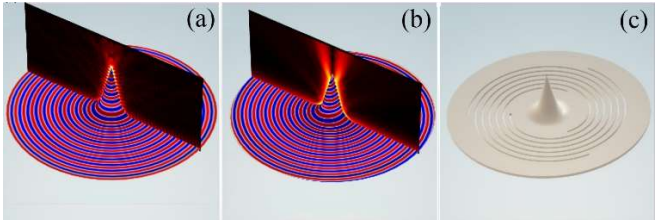

Fig. 1. (a), (b) Simulated electric field distribution for $l$ = 0 and for $l$ = 2 corresponding to the "trapping" and the "beaming" modes of the nanotip. (c) The schematic diagram of the 3D structure with an illustrative spiral grating with three arms.

The metallic cone can be treated as a cylinder with a varying radius $\rho(z)$ and its plasmonic eigenmodes then are given as [15,18]:

$$E_i(\rho,z) = \widetilde{E}_i(\rho, k_{r,i}) \exp(i\beta_l z) \exp(il\varphi) \qquad (1)$$

Here $\varphi$ is the azimuthal angle, $i = 1,2$ denotes the region outside or inside the metallic cylinder, $\beta_l$ is the complex propagation constant of the $l_{th}$ mode and $k_{r,i}$ is the transverse component of the wave vector given as $\epsilon_n k_0^2 = \beta_i^2 + k_{r,i}^2$, where $k_0$ is the vacuum wavenumber and $\epsilon_{1,2}$ are the dielectric constants of the vacuum and the metal respectively. The propagation constant, $\beta$ is found by solving the Helmholtz equation with a boundary conditions provided by the cylinder geometry. The dispersion of the $l_{th}$ plasmonic mode on a tapered cone obtained from Eq. (1) shows different behaviors for the $l$ = 0 and the $l$ > 0 modes [18]. For the $l$ = 0 mode (purely radial polarization) both the real and the imaginary parts of the $N_{eff}$ = β/k$_0$ tend to infinity while all the higher modes have a descending effective index curve. This means that the radial mode slows down towards the tip end while the other modes accelerate up to the radius where $N_{eff}$ = 1. At this point the mode is not confined to the cylinder anymore and it detaches from the surface of the cone. Here we the beaming occurs.

In order to generate a $l_{th}$ mode, i.e. to launch a PV with a desired OAM, one can use an axially symmetric SPP launching grating and/or modify the incident illumination's topology by the external spatial light modulator such as a liquid crystal based plates [19]. Here we use a specially designed spiral grating surrounding the tip [15]. Our SPP launching structure consists of spiral slits with the radii given by $R_m(\varphi) = R_0 + m \cdot \varphi/k_{SPP}$, where $R_0$ is the smallest radius of the slit and $m$ is the topological order of the spiral. The tip was located at the center of the spiral as shown in Fig. 1c. The base curvature and other geometric parameters were optimized to provide a smooth and the most efficient SPP propagation [15]. When this structure was illuminated from the bottom with the incident spin of $s_i = \pm 1$ (± corresponds to the right and left-handed state respectively) the spiral has generated a PV with a helical phase $exp(il\varphi)$. Due to the plasmonic spin-orbit coupling the topological charge of the SPP field is given by $l = m + s_i$. Therefore by properly designing the structure and the incident light state, the OAM of the PV propagating on the cone can take arbitrary unitary values.

We now investigate the dynamics of a nanoparticle that is free to move in a close vicinity of the proposed structure. We assume the refractive index and radius of the particles are $n_1$ = 1.45 and $R$=5 nm, respectively. When $R \leq \lambda_0/20$ ($\lambda_0$ is the vacuum wavelength), the Rayleigh approximation condition is satisfied, and the particle can be modeled as a simple dipole. In this case, the Rayleigh scattering theory can be used to calculate the optical trapping force [6,20–23]. When the particles are placed on the surface of the structure, they are subjected to three forces: gradient force, scattering force and gravity. In the simulation, the particle has a density of 2.648×10$^3$ kg/m$^2$. Through a precise calculation, we found the scattering force is almost 1/1000 of the gradient force and the gravity has even weaker effect on the particles motion. Therefore the gradient force plays the major role in controlling the particles' dynamics while the other two forces can be neglected. The gradient force can be expressed as [22]:

$$F_{grad}(r,z) = \frac{2\pi n_2 R^3}{c} \left(\frac{m^2-1}{m^2+2}\right) \nabla I(r,z) \qquad (2)$$

Here n$_2$ is the refractive index of surroundings medium, R is the particle radius, c is the speed of light, m = n$_1$/n$_2$, and $I$(r ,z) is the light intensity which can be given as:

$$I(r,z) = \frac{n_2 \varepsilon_0 c}{2} |E(r,z)|^2 \qquad (3)$$

Where $\varepsilon_0$ is the permittivity of free space, and E(r, z) is the electric field. It can be seen that the magnitude of gradient force is proportional to the gradient of electric field intensity.

As shown in Fig. 2, the force points to the position where the electric field reaches a maximum all the time. Significant differences are shown between the case with $l$ = 0 and with $l$ > 0. As mentioned before, in the first configuration $N_{eff}$ tends to infinity and consequently the EM field is confined at the tip as in the classical nanofocusing effect [24]. The light intensity gets largest at the tip of the cone, which leads to an expected trapping effect. For higher $l$ values, on the contrary, accordingly with the decoupling of the plasmonic vortex to light beam, the space above the tip generate a gradient force field able to constrain the trajectory of particles, forming a particle beam.

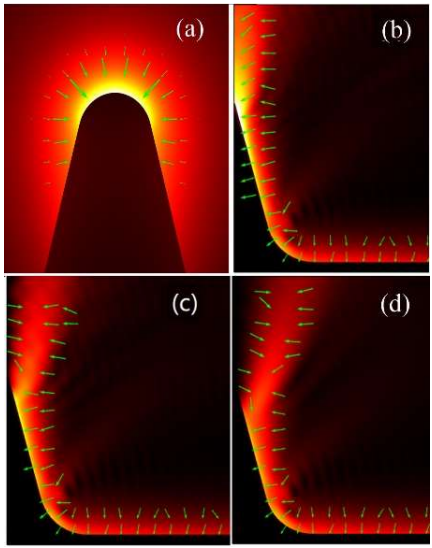

Fig. 2. The calculated gradient force. The arrows give the direction of the gradient force accompanying with the electric field. (a-d) Cases for $l$ = 0, 1, 2, 3.

To be vividly illustrated, we implemented numerical simulations of particle tracking considering that the nanoparticles are free to move in space around the structure. Fig. 3(a) illustrates the case where the particles are initially evenly placed near the tip of the cone at t = 0 and released from this position. At t=0.1 s, the gradient force, calculated as:

$$V(r,z) = -\pi n_2^2 \varepsilon_0 R^3 \left(\frac{m^2-1}{m^2+2}\right) | E(r,z)|^2, \qquad (4)$$

generates a potential well that is able to stably trap the nanoparticle, as illustrated in Fig. 3(b).

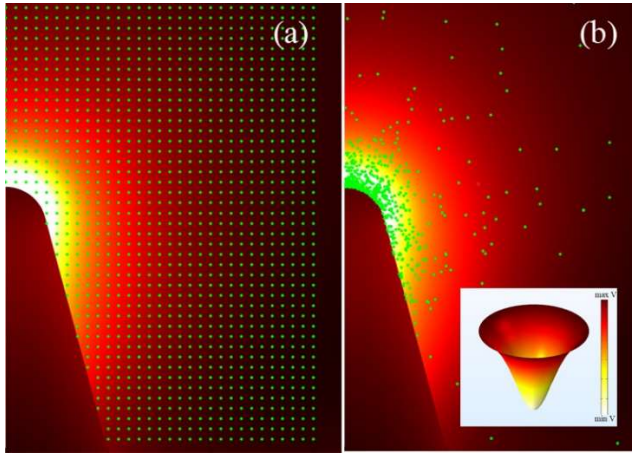

Fig. 3. Trapping case for topological charge $l$ = 0, tip's aperture of 30° and particles with refractive index $n_1$ = 1.45 and radius $R$ = 5 nm. (a) The particles are evenly placed near the tip of the cone when $t$ = 0. (b) Most of the particles are attracted to the tip of the cone when $t$ = 0.1 μs. (inset) Corresponding potential well.

By increasing the topological charge of the PV the particle tracking simulations, as expected, show a completely different behavior. As illustrated in Fig. 4, the nanoparticles placed 5 μm from the tip are always pushed by the PV along the shown trajectories. These trajectories depend upon the induced topological charge. This phenomenon stems from the effect that was observed in [15]. The nanotip, in fact, enables the decoupling of the PV to freely propagating light beam at a specific height of the tip, as illustrated in Fig. 4. The EM field generated by the structure is then able to control the motion of the nanoparticle accordingly. It is worth noting that this behavior is enabled by the optimized curvature radii at the tip base. This provides the condition for the smooth and reflection-less PV propagation along the tip, ensuring an almost perfect coupling of energy to the far field.

The observed phenomenon suggests a potential scheme of the dynamic control of the nanoparticle. In fact, by externally altering the topological charge of the PV the system can be tuned into different states suitable for trapping or releasing the nanoparticles. As was earlier mentioned, it is possible to switch from $l$ = 0 to higher $l$ values by using the proper combination of the coupler structure geometry and the impinging light spin. Moreover, by monitoring the dynamics of a nanoparticle interacting with the structure it can be also possible to evaluate the total momentum of the excitation light.

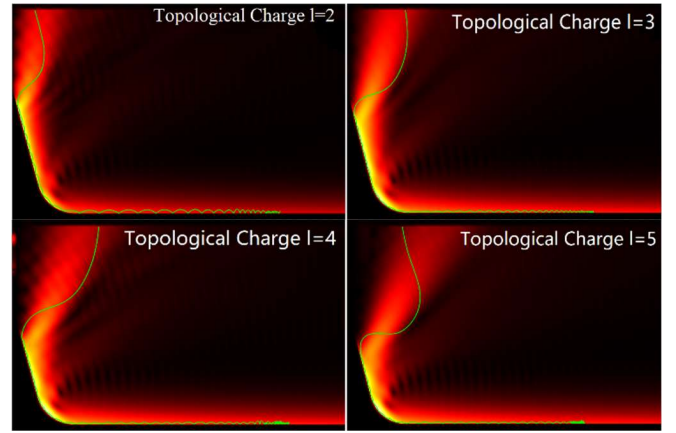

Fig. 4. Tweezing/Beaming case for high-order topological charge. The particles are statically released, 5 nm from the bottom, 10 μm away from the tip. Green lines are particle trajectories.

In conclusion, here we show how a specifically designed 3D nanostructure supporting PVs can be used to manipulate and trap dielectric nanoparticles. This is achieved via the unique system geometry enabling generating of plasmonic modes carrying OAM. Our results show that by changing the input light polarization, the system can be excited in different configurations suitable for trapping particles or pushing them out plane at specific points. Finally, considering the ability of OAM carrying beam to induce a torque to nano-objects, we expect that our system can also be investigated to study chiral and non-chiral particles in an easy dynamically controllable platform.

**Acknowledgment**. The authors acknowledge the financial support by the Ministry of Science, Space and Technology of Israel. NM acknowledges the financial support from the FEDER program (grant n. 2017-03-022-19 Lux-Ultra-Fast).